\documentclass[a4paper,11pt]{article}
\usepackage{pos}
\usepackage{hyperref}
\hypersetup{
    colorlinks=true,    
    linkcolor=blue,     
    urlcolor=cyan       
}

\title{Electroweak Transitions Involving Resonances}

\author*[a,b]{Luka Leskovec}

\affiliation[a]{ Faculty of Mathematics and Physics,\\
University of Ljubljana, Jadranska 19, 1000}

\affiliation[b]{Jozef Stefan Institute, \\
  Jamova 39, 1000 Ljubljana, Slovenia}

\emailAdd{luka.leskovec@ijs.si}

\abstract{The increasing importance of hadronic resonances in our understanding of the Standard Model is underscored by recent advancements in lattice Quantum Chromodynamics (QCD) calculations. We review recent developments, with a particular emphasis on electroweak transitions that result in two-hadron final states. Additionally, we present the finite-volume lattice QCD methodologies that are pivotal in such studies in the context of preliminary results from the $B \to \pi\pi\ell\bar{\nu}$.}

\FullConference{The 40th International Symposium on Lattice Field Theory (Lattice 2023)\\
July 31st - August 4th, 2023\\
Fermi National Accelerator Laboratory\\}

\begin{document}
\maketitle

\section{Introduction}

Electroweak transitions, mediated by the $W^{\pm}$ and $Z$ bosons and the photon $\gamma$, are fundamental in understanding particle behavior, especially in processes involving quark flavor or lepton number changes. These transitions have significantly advanced our comprehension of the Standard Model, shedding light on electroweak symmetry breaking and the origins of particle masses. Furthermore, they are pivotal in exploring CP violation, illuminating our understanding of the universe's matter-antimatter asymmetry. Electroweak transitions also serve as a critical tool for probing the limits of the Standard Model and searching for new physics by investigating rare decays and anomalies that might indicate new particles or interactions.\\

Lattice QCD has been instrumental in determining the elements of the CKM matrix \cite{FlavourLatticeAveragingGroupFLAG:2021npn}, a cornerstone achievement in the field. The collaborative efforts of the past 15 years, coupled with global experimental endeavors, have yielded precise determinations of key Standard Model parameters, for example $V_{ud}$, $V_{us}$, $V_{cd}$, $V_{cs}$, to a few percent accuracy. However, these efforts have predominantly focused on initial and final states comprising QCD-stable hadrons, neglecting hadronic resonances.\\

\begin{figure}[htb!] 
\centering 
\includegraphics[width=0.5\textwidth]{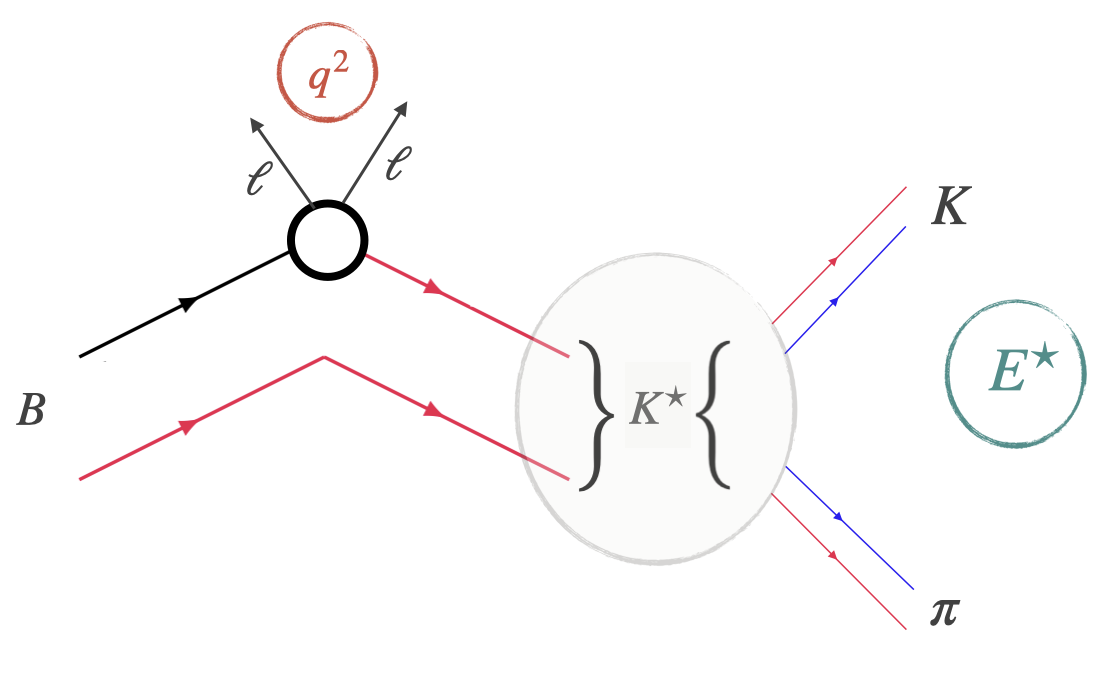} 
\caption{Illustrative representation of a rare $B$ decay, highlighting the transition of a $\bar{b}$-quark in the $B$-meson to a $s$-quark and a pair of leptons $\ell$, resulting in the formation of a $K^{\star}$ resonance which subsequently decays into a pion ($\pi$) and a kaon ($K$).}
\label{fig:rareB} 
\end{figure}

Hadronic resonances, a significant part of the hadron spectrum, are characterized by their brief lifetimes and rapid decay into other hadron states. In experiments, these resonances manifest as distinct enhancements in the energy-dependent cross-section of certain final states. A particularly interesting exemple is the rare decay of a $B$-meson, particularly the process $B\to K^\star(892) \ell \ell$, as depicted in Fig.~\ref{fig:rareB}.
The decay of the initial $B$-meson through the electroweak interaction results in a $\bar{b}$-quark transforming into a $\bar{s}$-quark and a pair of leptons $\ell$ via a loop process, as indicated in Fig.~\ref{fig:rareB}. The momentum transfer, denoted as $q^2$, corresponds to the energy carried away by the two leptons. The newly formed $\bar{s}$-quark, in combination with the existing light $u$-quark from the $B$-meson, generates a hadron with an $\bar{s}$-quark and a light $u$-quark. Due to its nature, this process allows for the creation of hadronic states with angular momentum $J=0, 1,\ldots$. Below the three-particle threshold, these states predominantly manifest as kaon ($K$) and pion ($\pi$) pairs at a specific two-hadron invariant mass, $E^\star$. The dynamic interplay of $K$ and $\pi$, which are stable against strong interaction decay, enters in their rescattering process, which can induce the formation of a $K^\star$ resonance, an unstable hadron state observed as a significant enhancement in the $E^\star$ dependence of the partial cross-section in $K\pi$ scattering \cite{Wehle:2016gfb}.\\

While $K^\star$ resonances are often treated as stable hadrons for analytical simplicity, it is essential to recognize their highly transient nature, with lifespans of approximately $10^{-23}$ seconds before decaying through Quantum Chromodynamics (QCD) into a $K$ and $\pi$ pair \cite{ParticleDataGroup:2022pth}. Therefore, the relevant final states extend beyond the ephemeral $K^\star$ resonances to the QCD asymptotic states, particularly the $K$ and $\pi$. To effectively discuss Electroweak Transitions Involving Resonances, it is necessary to consider Electroweak Transitions leading to multihadron final states; more specifically two-hadron final states. This text is organized as follows: Sec.\ref{sec:2} reviews the most significant processes in lattice QCD and experiment. Sec.\ref{sec:3} provides an overview of the lattice QCD methodologies applied in these calculations, along with an example workflow. Sec.~\ref{sec:4} discusses some promising topics for the near-term future, and Sec.\ref{sec:Summary} offers a summary.

\section{Current Status} \label{sec:2}

We consider three types of electroweak processes involving multi-hadron final states: purely hadronic electroweak transitions, photoproduction of hadronic resonances, and semileptonic electroweak processes.

\paragraph{Purely Hadronic Electroweak Transitions}
The primary focus within purely hadronic electroweak transitions is the $K\to \pi\pi$ process, critical for enhancing our understanding of CP violation and the pronounced matter-antimatter asymmetry, which diverges significantly from Standard Model predictions. In this process, an $s$-quark in the Kaon emits a $W$ boson, converting it into a $u$ quark. The $W$ boson, rather than decaying into a lepton-antineutrino pair, produces a pair of $u$ and $d$ quarks. These quarks subsequently form hadrons with the remaining light quarks, resulting in two pions in the final state. These pions can exist in two isospin states, $I=0$ and $I=2$, with the relevant partial waves being $\ell=0$ and $\ell=2$ due to Bose symmetry.

\begin{figure}[htb!]
    \centering
    \includegraphics[width=0.5\textwidth]{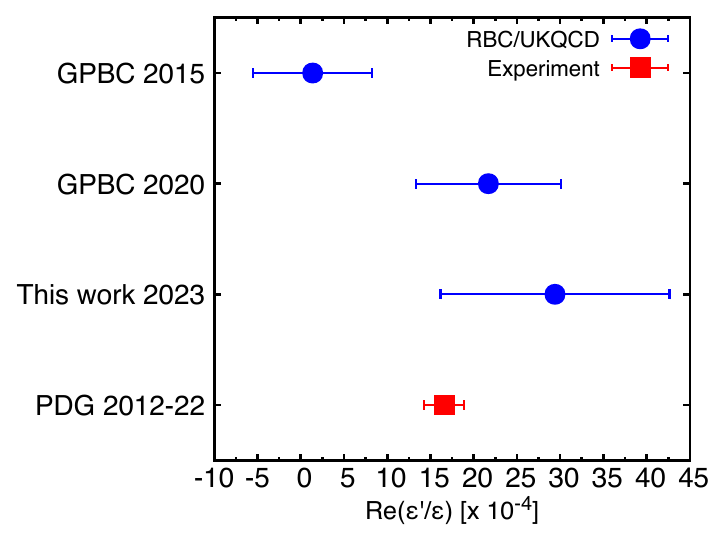}
    \caption{Figure taken from Ref.~\cite{RBC:2023ynh}; shown is the value for $Re(\tfrac{\epsilon'}{\epsilon})$ from RBC/UKQCD calculations as well as the PDG value.}
    \label{fig:Kpipi}
\end{figure}

Three lattice QCD calculations of the $K \to \pi\pi$ process exist: two by the RBC/UKQCD collaboration \cite{RBC:2015gro,RBC:2020kdj,RBC:2023ynh} and one by Ishizuka et al.~\cite{Ishizuka:2018qbn}. Both RBC/UKQCD calculations use M\"obius Domain wall fermions at quark masses corresponding to the physical point; they, however, employ different lattice spacings: Ref.~\cite{RBC:2015gro,RBC:2020kdj} used $a=0.1433$ fm and Ref.~\cite{RBC:2023ynh} $a\approx 0.1932$ fm. The first calculation \cite{RBC:2015gro} and its improvement \cite{RBC:2020kdj} made use of the twisted-G-parity\footnote{A G-parity condition involves transforming the quark-lines with Charge conjugation combined with an isospin rotation about the $y$-axis} boundary condition to maximize the signal, which is best when the two pions final state energy matches the initial state kaon energy. The second RBC/UKQCD calculation \cite{RBC:2023ynh} made use of distillation \cite{HadronSpectrum:2009krc} combined with periodic boundary conditions to construct the initial and final states. The crucial part of both calculations was a well-determined scattering phase shift for the $I=0, \ell=0$ $\pi\pi$ channel. The resulting $Re(\tfrac{\epsilon'}{\epsilon})$ values are shown in Fig.~\ref{fig:Kpipi}, and the reader is directed to Ref.~\cite{RBC:2023ynh} for further details. The calculation by Ishizuka et al.~\cite{Ishizuka:2018qbn} was performed with Clover fermions with quark masses corresponding to $m_\pi\approx 260$ MeV and a single lattice spacing of $0.091$ fm and found a ratio of $Re(\tfrac{\epsilon'}{\epsilon})=1.9(5.7)\times 10^-3$; At the same time, their result is still statistically compatible with zero (and thus not shown in Fig.~\ref{fig:Kpipi}), further efforts are warranted to verify the $Re(\tfrac{\epsilon'}{\epsilon})$ results from RBC/UKQCD.

Experimentally, the $K\to\pi\pi$ processes were measured at the NA48 \cite{NA48:2001bct} and KTeV experiments \cite{Worcester:2009qt}. They remain a primary source of background in ongoing experiments like KOTO \cite{Aoki:2021cqa,Nanjo:2023xvj} and NA62 \cite{NA62:2020upd}, necessitating a robust Standard Model understanding. Hence, precise lattice QCD determinations of these processes continue to play a crucial role in our comprehension of the Standard Model.\\

In the community's effort to understand CP violation, Hansen et al. are pushing forward our understanding of CP violation in the charm sector with pioneering calculations of the $D \to \pi \pi$ process \cite{hansen:latt2023}. At the same time, there are plenty of other processes of interest; they are more challenging from both the formal and computational approaches.

\paragraph{Photoproduction of Hadronic Resonances}
In the realm of photoproduction of hadronic resonances, lattice QCD calculations have been conducted for the $\pi\gamma \to \pi\pi$ and $K\gamma \to K\pi$ processes, while a toy model analysis discusses the coupled channel approach and preliminary results of the $\gamma \to \pi\pi \vert K \bar{K}$ demonstrate the feasibility of a coupled channel calculation. The $\pi \gamma \to \pi \pi$ process, besides serving as a testbed for lattice QCD methodologies, contributes to determining the dispersive determination of the $(g-2)\mu$ by ascertaining the chiral anomaly $F{3\pi}$, and  $\gamma \to \pi\pi \vert K \bar{K}$ might offer insight into where the pure Yang-Mills glueballs have gone in the dynamic QCD spectrum.\\

The $\pi \gamma \to \pi \pi$ process is perhaps the simplest for testing the relevant lattice QCD methodologies. There, the photon excites the pion ($I(J^P)=1(0^-))$, where the spins of the quarks are anti-parallel, to a $\rho$ $(I(J^P)=1(1^-))$, where the spins are parallel. As the $\rho$ is a hadronic resonance, it decays to a pair of pions in $P$-wave with a branching fraction larger than $99.9$ \% and a decay width of approximately $\Gamma = 150$ MeV \cite{ParticleDataGroup:2022pth}. The first lattice QCD determination of the $\pi \gamma \to \pi\pi$ transition amplitude was performed by HadSpec \cite{Briceno:2016kkp}, where they used Clover-Wilson fermions on an anisotropic lattice with quark masses corresponding to $m_\pi \approx 400$ MeV and a single lattice spacing, $a_s\approx 0.12$ fm. LHPC followed up the calculation \cite{Alexandrou:2018jbt} on isotropic lattices with Clover-Wilson fermions and a pion mass of $\approx 320$ MeV and a lattice spacing of approximately $0.114$ fm. The combined results were used in the determination of $F_{3\pi}$ in Ref.~\cite{Niehus:2021iin}.\\

The $K\gamma \to K\pi$ process is very similar to the $\pi \gamma \to \pi\pi$, except an $ s$-quark replaces a light quark. HadSpec conducted the sole calculation for this process. So far, the only calculation of this process was determined by HadSpec \cite{Radhakrishnan:2022ubg}, where they used Clover-Wilson fermions on a single gauge ensemble with a lattice spacing $a_s\approx 0.12$ fm and a light quark mass corresponding to $m_\pi \approx 280$ MeV.\\

The $\gamma \to \pi\pi|K\bar{K}$ process produces states out of the vacuum; in this case, it creates a pair of pions or a pair of kaons. The two pions can resonate and produce a $\rho$ resonance, while the $K\bar{K}$ seems mostly decoupled \cite{Wilson:2015dqa}. The time-like pion form factor can be determined from the transition amplitude, which plays an important role in the $(g-2)_\mu$ \cite{Meyer:2011um}; the process also offers insights into the quark content of hadronic resonances and has implications for the understanding of glueballs outside pure Yang-Mills theory \cite{Briceno:2015csa}.\\

Experimentally, the photoproduction (or the reverse, radiative decay) processes are measured at the GlueX \cite{Accardi:2023chb} experiment at JLab and (hopefully one day) also at PANDA at GSI \cite{PANDA:2021ozp}. 

\begin{figure}[htb!]
    \centering
    \includegraphics[width=0.55\textwidth]{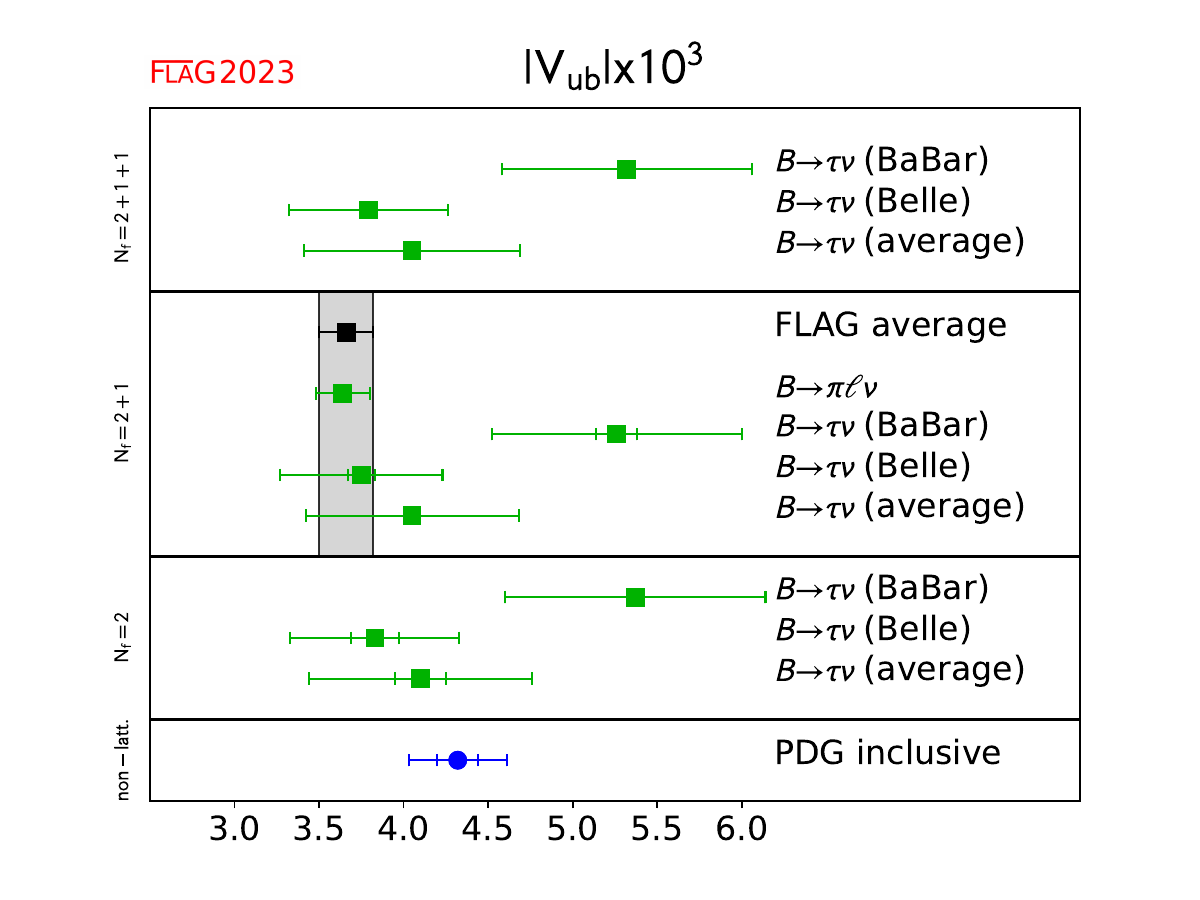}
    \caption{Figure taken from Ref.~\cite{FlavourLatticeAveragingGroupFLAG:2021npn}; shown are the $V_{ub}$ matrix elements as determined from different experiments and different processes; both inclusive and exclusive analyses are shown. The average value its uncertainty for $V_{ub}=3.64(16)$ \cite{FlavourLatticeAveragingGroupFLAG:2021npn} are dominated by the semileptonic decay, $B\to\pi\ell\bar{\nu}$.}
    \label{fig:flagVub}
\end{figure}

\paragraph{Semileptonic Electroweak Processes}
Although no published works currently exist on semileptonic electroweak processes involving resonances with a complete finite-volume treatment, preliminary results on the $B \to \pi \pi \ell \bar{\nu}$ process are discussed in Sec.~\ref{sec:3}. In this process, the $b$-quark in the $B$-meson emits a $W$ boson, which then further decays into a pair of $\ell \bar{\nu}$, while the $b$-quark changes to a $u$-quark. This transition offers a new probe for the smallest and least known of the CKM matrix elements, $V_{ub}$; the current value predominantly comes from the $B\to\pi\ell\bar{\nu}$ process, as can be seen in Fig.~\ref{fig:flagVub}.
Due to the large variation in the $V_{ub}$ determinations, and in particular due to the fairly large difference ($\approx 1.7 \sigma$ when all uncertainties are included \cite{FlavourLatticeAveragingGroupFLAG:2021npn}) between the inclusive and exclusive determination, an additional, new exclusive process to determine $V_{ub}$ is needed to understand the source of the tension better. The process $B\to \pi\pi (\to \rho) \ell \bar{\nu}$, actively measured at Belle II \cite{Belle-II:2022fsw} and Belle \cite{Belle:2020xgu}, serves as a crucial alternative for determining $V_{ub}$, helping resolve tensions between inclusive and exclusive determinations. Furthermore, as it has more form factors than the $B\to\pi\ell\bar{\nu}$ process, it is also better for searching for potential Beyond the Standard Model contributions.\\

Semileptonic decays extend beyond mesons to baryons, particularly in the radiative and electroweak productions of the $\Delta(1232)$ resonance on a nucleon, a process vital for the success of the DUNE experiment \cite{NuSTEC:2017hzk,Ruso:2022qes,Simons:2022ltq}.

\section{Lattice Technology} \label{sec:3}

\begin{figure}[htb!]
    \centering
    \includegraphics[width=0.99\textwidth]{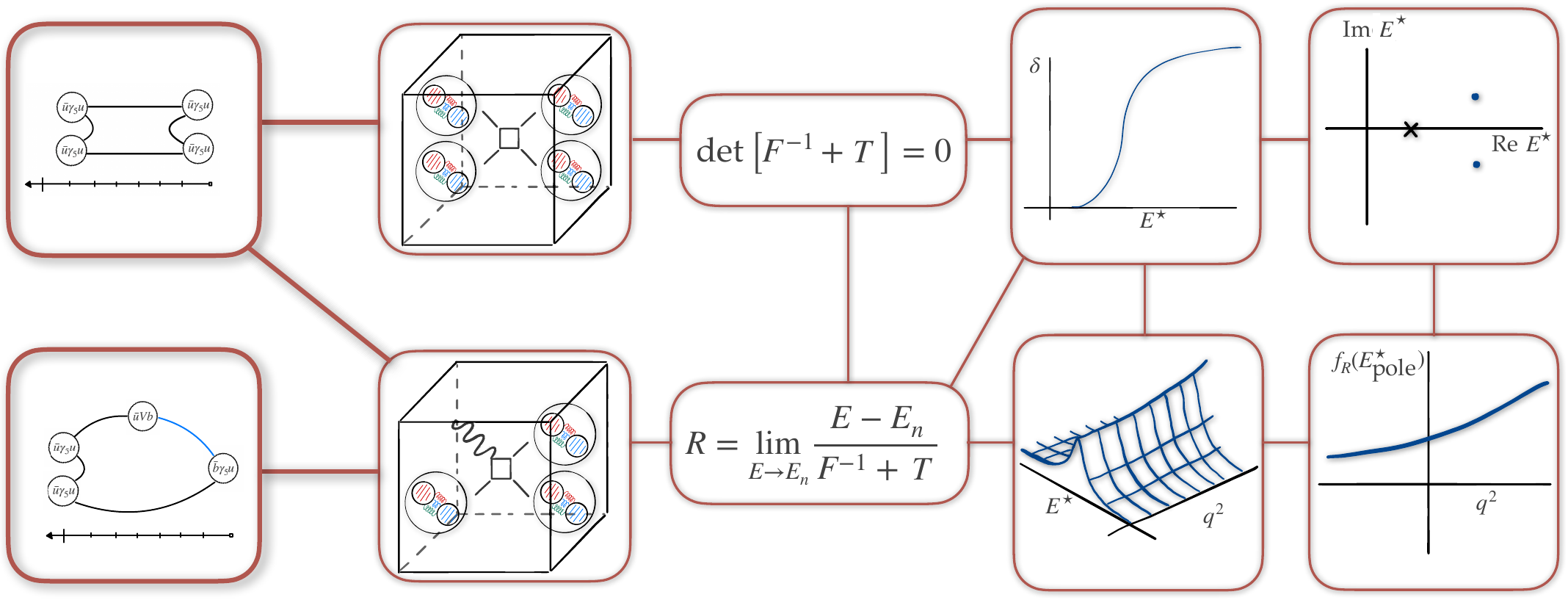}
    \caption{Schematic representation of the process to determine infinite-volume scattering and transition amplitudes from lattice QCD data. The top half illustrates the spectrum analysis workflow, while the bottom details the matrix element analysis.}
    \label{fig:scheme}
\end{figure}

In lattice QCD, the determination of infinite-volume transition amplitudes, as depicted in Fig.~\ref{fig:scheme}, involves a comprehensive workflow encompassing both spectrum and matrix element analysis: the top half of the diagram schematically shows the spectrum analysis, while the bottom half shows the matrix element analysis.\\

The spectrum analysis, now a routine computation in many groups, starts with defining the energy region of interest. For the $\rho$ resonance, it is sufficient to go approximately to the $K\bar{K}$ threshold\footnote{As the $K\bar{K}$ has a minimal coupling to the $\rho$ resonance \cite{Wilson:2015dqa}.}. We pick a set of interpolating operators $\mathcal{O}_a$ of the necessary types - for the $\rho$ on our ensemble with $m_\pi\approx 320$ MeV and $a\approx 0.114$ fm, two types suffice: $\bar{q}q(P)$ and $\pi(p_1)\pi(p_2)$, where $P$ is the size of the total momentum of the system and $p_1$ and $p_2$ are the sizes of the momenta belonging to the two pions. They are connected to $P$ through $\vec{P}=\vec{p}_1+\vec{p}_2$. From these interpolating operators, we construct a correlation matrix $C_{ab}(t) = \langle 0 | \mathcal{O}_a(t) \mathcal{O}^\dagger_b(0)  | 0 \rangle$, which we proceed to calculate through Wick contractions and whichever technology we prefer to employ for the propagators - this is represented by the left-most diagram on the top half of Fig.~\ref{fig:scheme}. In the preliminary results presented below, we employ a combination of forward, sequential, and stochastic propagators \cite{Alexandrou:2017mpi}. However, distillation is a great alternative when complicated sources are needed \cite{HadronSpectrum:2009krc}.\\

The correlation matrix $C_{ab}(t)$ contains information on the spectrum $E_n$ and the overlap between the states, $n$, in the energy region and the interpolating operators, $a$, $Z_a^n$:
\begin{align}
\label{eq:C}
    C_{ab}(t) = \sum_n \frac{Z_a^nZ_b^{n,*}}{2 E_n} e^{-E_n t}.
\end{align}
To extract the overlaps and the spectrum from the correlation matrix, we solve the variational problem:
\begin{align}
\label{eq:GEVP}
    &C_{ab}(t) u_b^n(t) = \lambda_n(t,t_0) C_{ab}(t_0) u_b^n(t), \cr
    &u_a^n(t) C_{ab}(t) u_b^m(t) = \delta^{nm},
\end{align}
Where $u$'s are the generalized eigenvectors and play an important role also in the matrix element analysis; $\lambda_n(t,t_0)$ are principal correlators, which can be fit with a one, two, or more state models to determine the spectrum $E_n$. We can repeat this analysis on multiple volumes and in multiple irreducible representations. For the presented data, we employ a single volume, $32^3 \times 96$, with a lattice spacing of $a=0.114$ fm and all irreducible representations that have a $J^P=1^-$ state in them for total momenta $P<\frac{2\pi}{L}\sqrt{3}$ \cite{Alexandrou:2017mpi}.

The spectrum, i.e., the set of energies $E_n$, can be related to the infinite-volume scattering amplitude $T$ in what is often referred to as the L\"uscher analysis\footnote{Sometimes referred to also as the finite-volume analysis.}, as depicted in the second-to-left and middle top diagram of Fig.~\ref{fig:scheme}. The Quantization Condition was first derived by L\"uscher \cite{Luscher:1990ux} in the context of relativistic quantum mechanics:
\begin{align}
    \det \left[ F^{-1}(E) + T(E) \right]\bigg\vert_{E=E_n} = 0,
\end{align}
followed up by Ref.~\cite{Kim:2005gf} with a Quantum Field Theory derivation, where in a very simplistic view, the finite-volume two-point correlation function as a function energy $E$, instead of time $t$ as in Eq.~\ref{eq:C}, is expanded as a sum over poles; each pole belongs to a particular energy level. Its position is on the real axis and depends on the lattice parameters, $L$ and quark masses, and the scattering amplitude $T$. Further information about the L\"uscher analysis can be found in Refs.~\cite{Rummukainen:1995vs,Leskovec:2012gb,Briceno:2014oea,Briceno:2017max,Woss:2020cmp}.\\

At this point of the analysis, there are already two parts that involve fitting data to mathematical models, which can, in principle, introduce changes in the data - to quantify this uncertainty, we vary the mathematical models we fit, both for the principal correlators as well as the scattering amplitude and assign a systematic uncertainty to the parameters.\\

The resulting scattering amplitude can then be used to calculate the cross-section, although not very useful, as we have yet to learn how to make an experiment that will scatter two pions. We can, however, calculate the phase shift $\delta$ - first-from-middle diagram of Fig.~\ref{fig:scheme}, which is related to the scattering amplitude $T\propto \frac{1}{\cot{\delta}-i}$. If we are looking for hadronic resonances (and bound/virtual bound states included), we search for poles of the scattering amplitude $T$ in all the relevant Riemann sheets.\\

\begin{figure}
    \centering
    \includegraphics[width=0.3\textwidth]{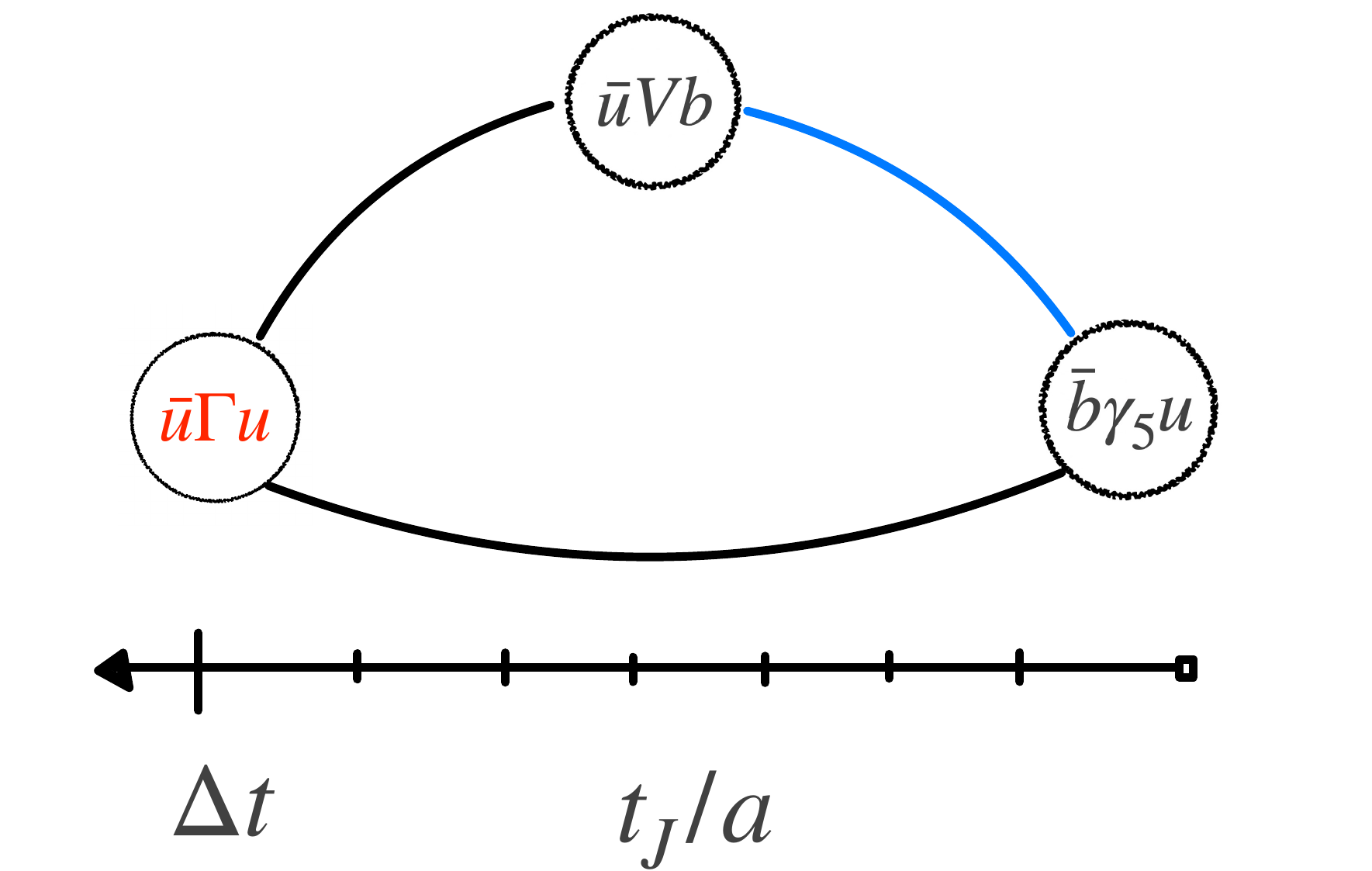}%
    \includegraphics[width=0.3\textwidth]{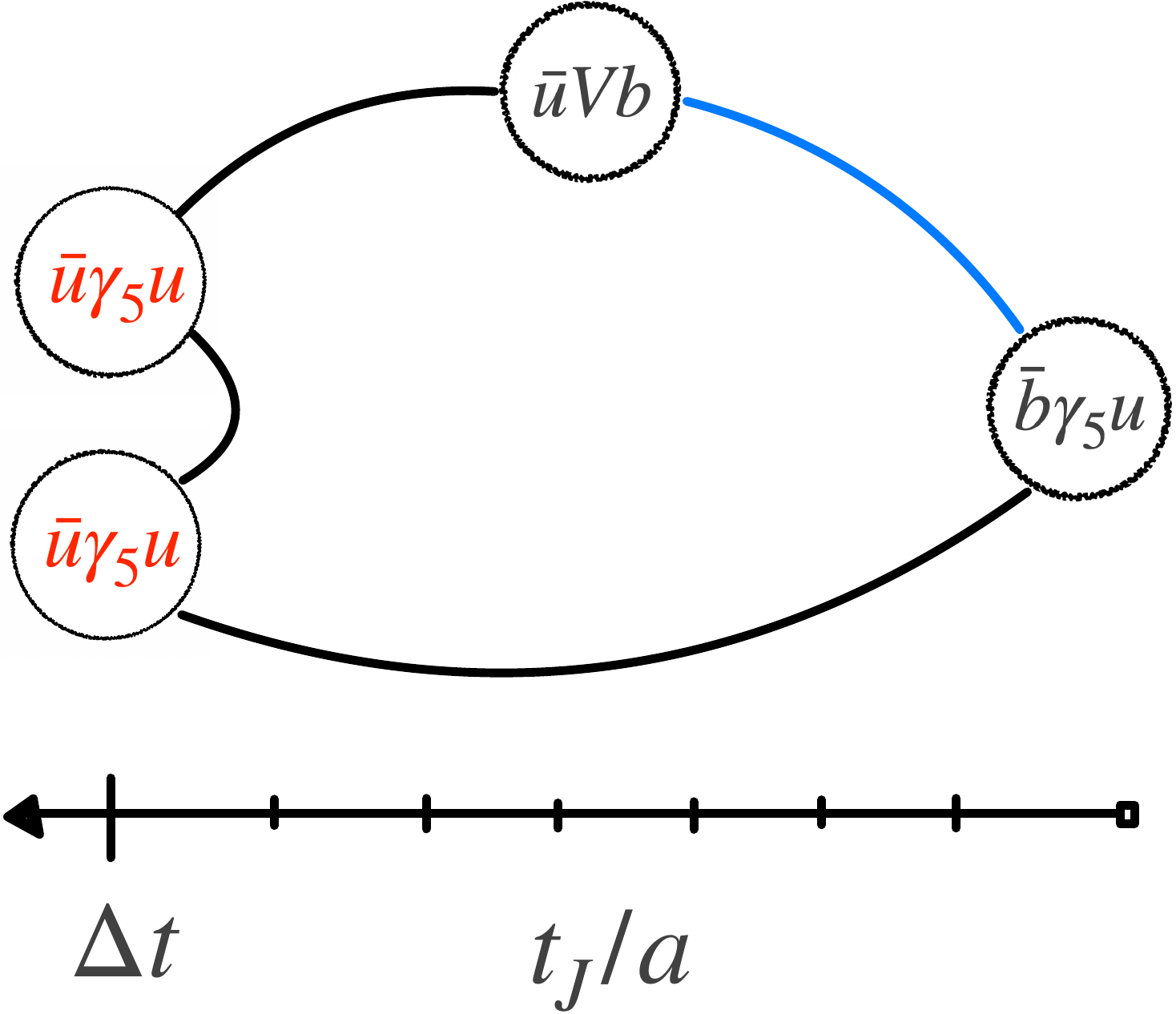}
    \caption{Wick contractions for the $B\to\pi\pi\ell\bar{\nu}$ process. The $B$-meson is located at $t=0$, the interpolating operator for the $\rho$, $\bar{u}\Gamma u$, is located at $\Delta t$ and the current insertion is located at $t_J$.}
    \label{fig:Wick3}
\end{figure}

The matrix element analysis is similar in spirit - except that now we have a new process in mind. The $B\to\pi\pi\ell\bar{\nu}$, has a $B$-meson at the source, which propagates forward until the $b$-quark transitions to a $u$-quark through an emission of a $W$ boson; the transition is taken into account as an external local current insertion, $J_{EW}=\bar{u}(\gamma_\mu - \gamma_5\gamma_\mu) b$. One option after the current is that the light quark from the current connects with the light quark from the $B$-meson, and another option is that a pair of light quark-antiquarks are made up from the vacuum, and the light quark connects with one of them. The latter is described by an interpolating operator $\mathcal{O}_a$, which consists of the same types of operators as in the spectroscopy analysis. The three-point correlation functions then look like:
\begin{align}
    \label{eq:C3}
    C_{3}^a = \langle 0 |\mathcal{O}_a(\Delta t) J^\mu(t_J) \mathcal{O}_B^{\dagger}(0)| 0 \rangle,
\end{align}
where $\mathcal{O}_B$ is the $B$-meson interpolator, and $J^\mu = \sqrt{Z^{u}Z^{b}} (\bar{u}\Gamma b + d^{(b)} \bar{u} \Gamma \gamma^i \nabla_i b)$ is the $O(a)$ improved current insertion. In the current, $\Gamma$ is either $\gamma^\mu$ for the vector current and $\gamma^\mu\gamma_5$ for the axial vector current; $d^{(b)}$ is the improvement coefficient, and $Z^{f}$ are renormalization coefficients of the flavor-conserving temporal vector current for quark flavor $f=u,b$. The three-point functions are computed from the Wick contractions shown in Fig.~\ref{fig:Wick3}; in the preliminary results shown here, we use a combination of forward and sequential propagators to construct them, although, as for the spectroscopy part, distillation can be used instead.
The three-point function contains the information about the matrix elements, $\langle \pi\pi, n| J_\mu | B \rangle$:
\begin{align}
    \label{eq:C3decomp}
    C_{3}^a  = \sum_{n \in \pi\pi} \sum_{m \in B}  &Z_a^n \; \langle n |J_\mu | m \rangle \; Z_B^m \; \frac{e^{-E_n (\Delta t - t_J)} e^{-E_m (t_J )}}{2E_n 2E_m},
\end{align}
the sum $n \in \pi\pi$ is the sum over all the states on the final state, and the sum $m \in B$ is the sum over all the states in the initial state. We are not interested in the excited states in the initial state. As the $B$-meson interpolator describes the ground state well, we can forget the sum over $m \in B$. 
\begin{figure}[htb!]
    \centering
    \includegraphics[width=0.7\textwidth]{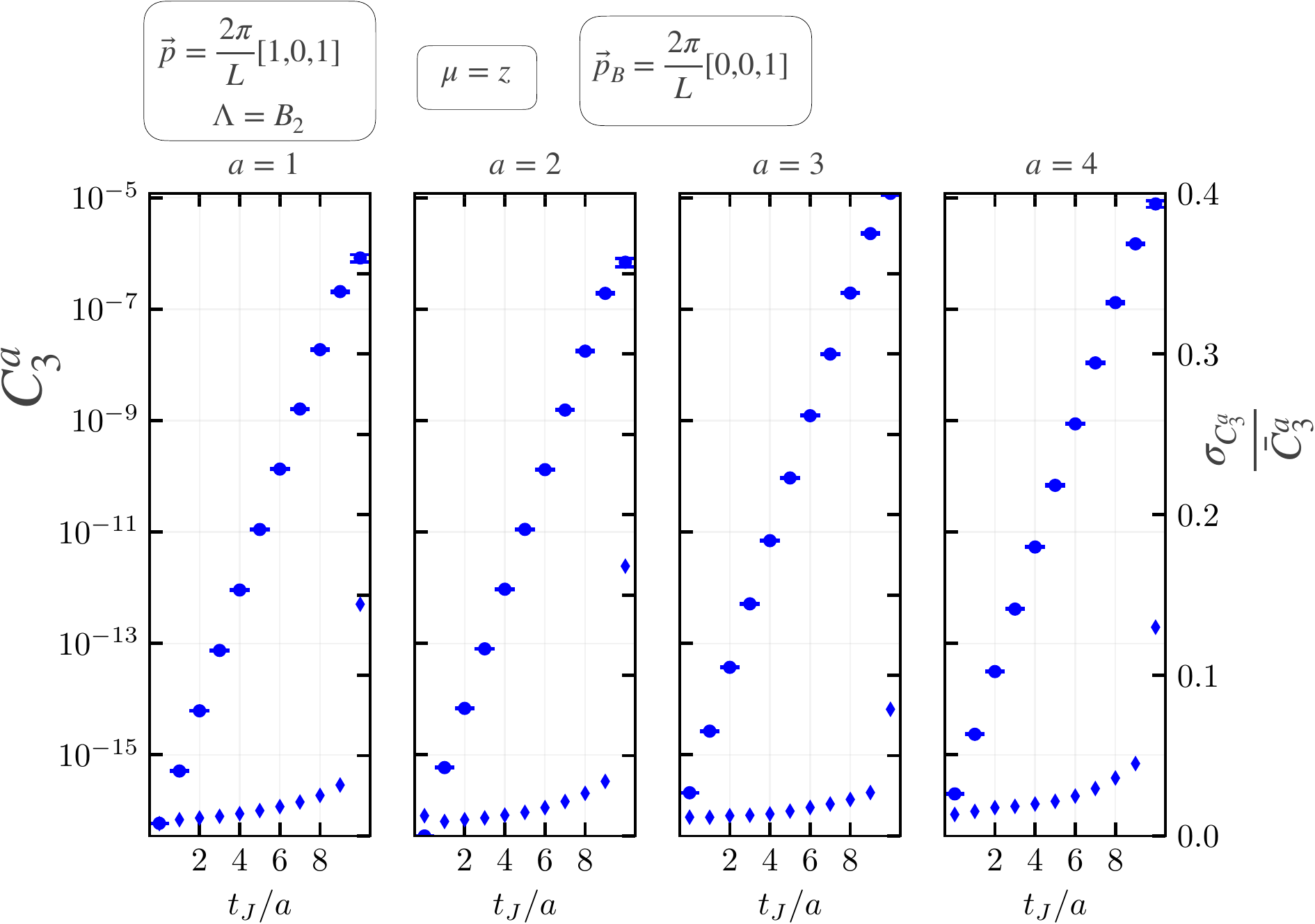}
    \caption{Example three point correlation functions in the irreducible representation $B_2$ of the $\vec{P}=\frac{2\pi}{L}(1,0,1)$, with $\vec{p}_B=\frac{2\pi}{L}(0,0,1)$. The circles with uncertainties represent the data, while the diamonds show the relative uncertainty of the data.}
    \label{fig:exampleC3}
\end{figure}
An example of the three-point correlation functions is shown in Fig.~\ref{fig:exampleC3}, where we show the irreducible representation $B_2$ of the $\vec{P}=\frac{2\pi}{L}(1,0,1)$ and a set of four interpolating operators:
\begin{align}
    \label{eq:operators}
    O_1^{B_2} &= \bar{q} \Gamma_{B_2} q, \cr
    O_2^{B_2} &= \bar{q} \gamma_t \Gamma_{B_2} q, \cr
    O_3^{B_2} &= \pi(p_1)\pi(p_2), \text{ and } \cr
    O_4^{B_2} &= \pi(p_1')\pi(p_2').
\end{align}
These three-point correlation functions still contain the sum over $n$ on the final state - on this side of the three-point correlation function, we also want to get excited state information. To achieve this, we make use of the results from the variational analysis - we construct optimized three-point functions according to \cite{Dudek:2009kk,Becirevic:2014rda,Shultz:2015pfa}:
\begin{align}
\label{eq:omega}
    \Omega_3^n = u_a^n C_3^a,
\end{align}
where we use one of the crucial aspects of the variational analysis. It allows us to build orthogonal states within our basis, reducing the sum over all the states $n \in \pi\pi$ to a specific state $n$. By using the relation:
\begin{align}
 u_a^n Z_a^m = \sqrt{2 E_n} e^{E_n t_0/2} \delta_{nm},
\end{align}
the sum in Eq.~\ref{eq:omega} reduces to:
\begin{align}
    \Omega_3^n = \sqrt{2 E_n} e^{E_n t_0/2} \; \langle n |J_\mu | B \rangle \; Z_B \; \frac{e^{-E_n (\Delta t - t_J)} e^{-E_B (t_J )}}{2E_n 2E_B}.
\end{align}
By simply removing the temporal dependence, we can determine the matrix elements $\langle n |J_\mu | B \rangle$ between the initial $B$-meson and the final, $n$-th state in the set of final volume $\pi\pi$ states. Alternatively we can construct a ratio \cite{Alexandrou:2018jbt},
\begin{align}
    R_n = \frac{\Omega_3^n\bar{\Omega}_3^m}{C_{B}(\Delta t)\lambda_n(\Delta t,t_0)},
\end{align}
where the $\bar{\Omega_3^n}$ three-point function has the quark-flows reversed, i.e. is the complex-conjugate of $\Omega_3^n$. These matrix elements would correspond to the second-to-left-most diagram in bottom half of Fig.~\ref{fig:scheme}.\\

\begin{figure}[htb!]
    \centering
    \includegraphics[width=0.95\textwidth]{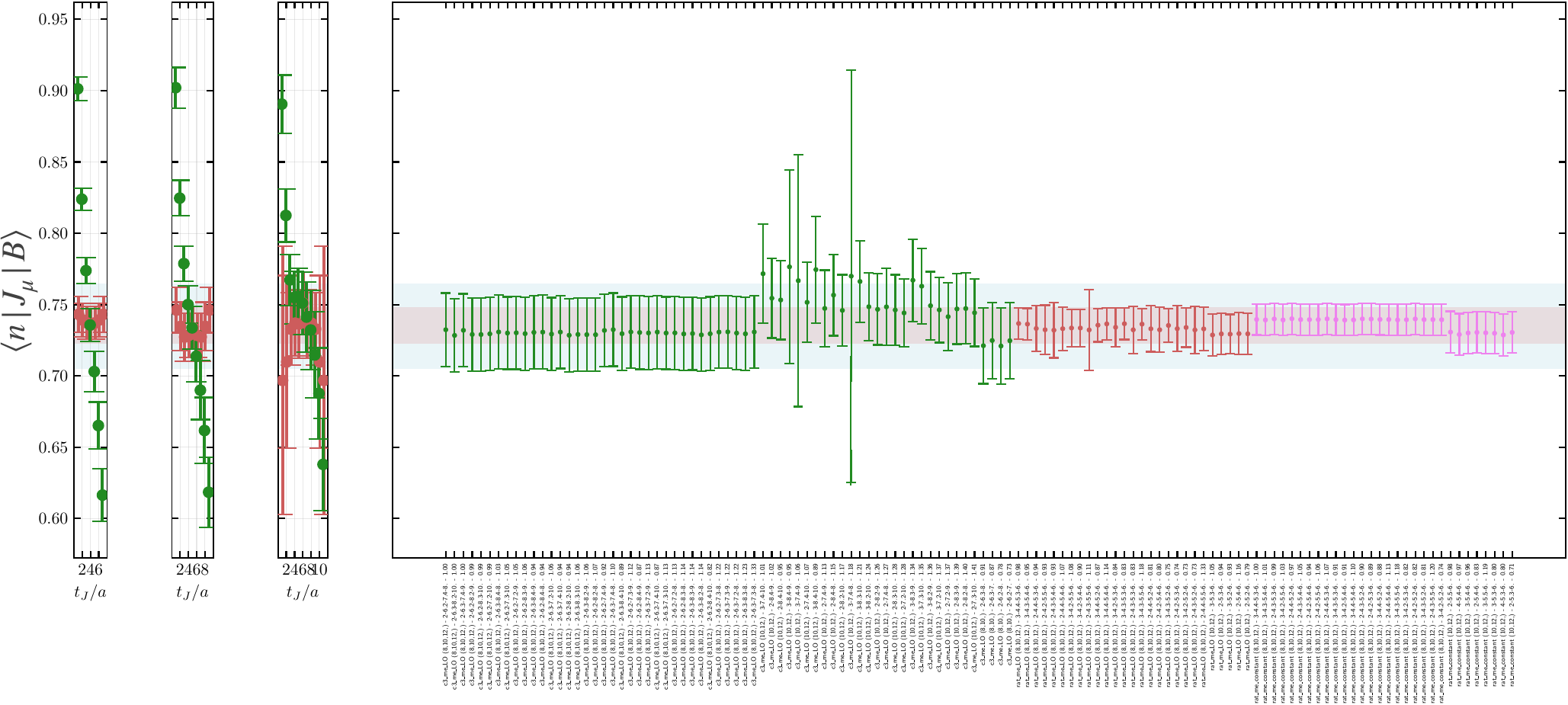}
    \caption{The Akaike Information Criterion matrix element average over all the models. The green-shaded region shows the matrix elements determined from the $\Omega_3^n$, and the red-shaded region shows the matrix elements determined from the $R_n$.}
    \label{fig:AIC}
\end{figure}

As is the case for the spectroscopy, so is the case for the matrix element analysis - when fitting, we make assumptions of the models, i.e., one-state fits, two-state fits, either on the initial state side or the final state side, as well as fitting both $\Omega_3^n$ and $R_n$ we produce a large number of fits. We employ the Akaike Information Criterion as outlined in Ref.~\cite{Jay:2020jkz} to average over the matrix elements determined in this way properly. In this manner, we produce matrix elements at a specific kinematical point of $E_n^\star$ and $q^2$, as seen in Fig.~\ref{fig:AIC}.
Two values for the matrix elements are determined this way - the green-shaded region, fitted to the $\Omega_3^n$ data, corresponds to our chosen fit. In contrast, the red-shaded region, fitted to the $R_n$ data, corresponds to our systematic fit. In this manner, we include a systematic uncertainty in our determination of the transition amplitude.\\

The matrix elements determined on the lattice are affected by finite-volume effects - an effect first pointed out by Lellouch and L\"uscher in their seminal work \cite{Lellouch:2000pv}, where they derived the normalization factor for the  $K\to\pi\pi$ matrix element. The calculation and the effects of the Lellouch-L\"uscher normalization factor have been discussed in large detail in Ref.~\cite{Lin:2001ek} and references therein, however the primary strategy was to tune kinematics (volumes and pion masses) to maximize the size of the matrix element. In Ref.~\cite{Briceno:2014uqa} Brice\~no, Hansen, and Walker-Loud derived the finite-volume effects from a Quantum Field Theoretical approach; the generalization changed the approach from tuning kinematics to using every possible state in the analysis.\\

\begin{figure}
    \centering
    \includegraphics[width=0.7\textwidth]{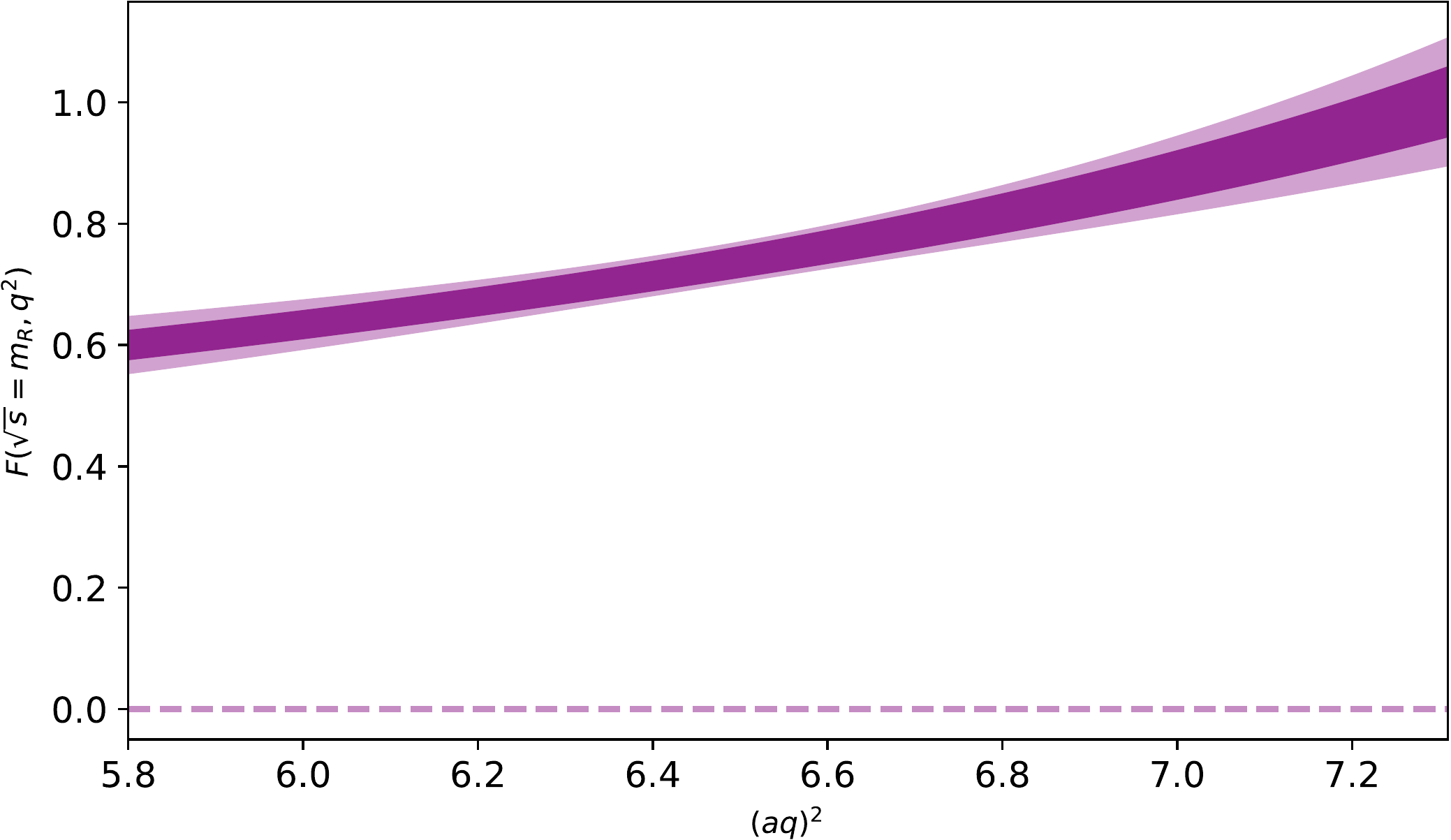}
    \caption{The average of all form factors; the dark-shaded region represents the statistical uncertainty, while the light-shaded region represents the systematical uncertainty.}
    \label{fig:ffx}
\end{figure}

\begin{figure}
    \centering
    \includegraphics[width=0.8\textwidth]{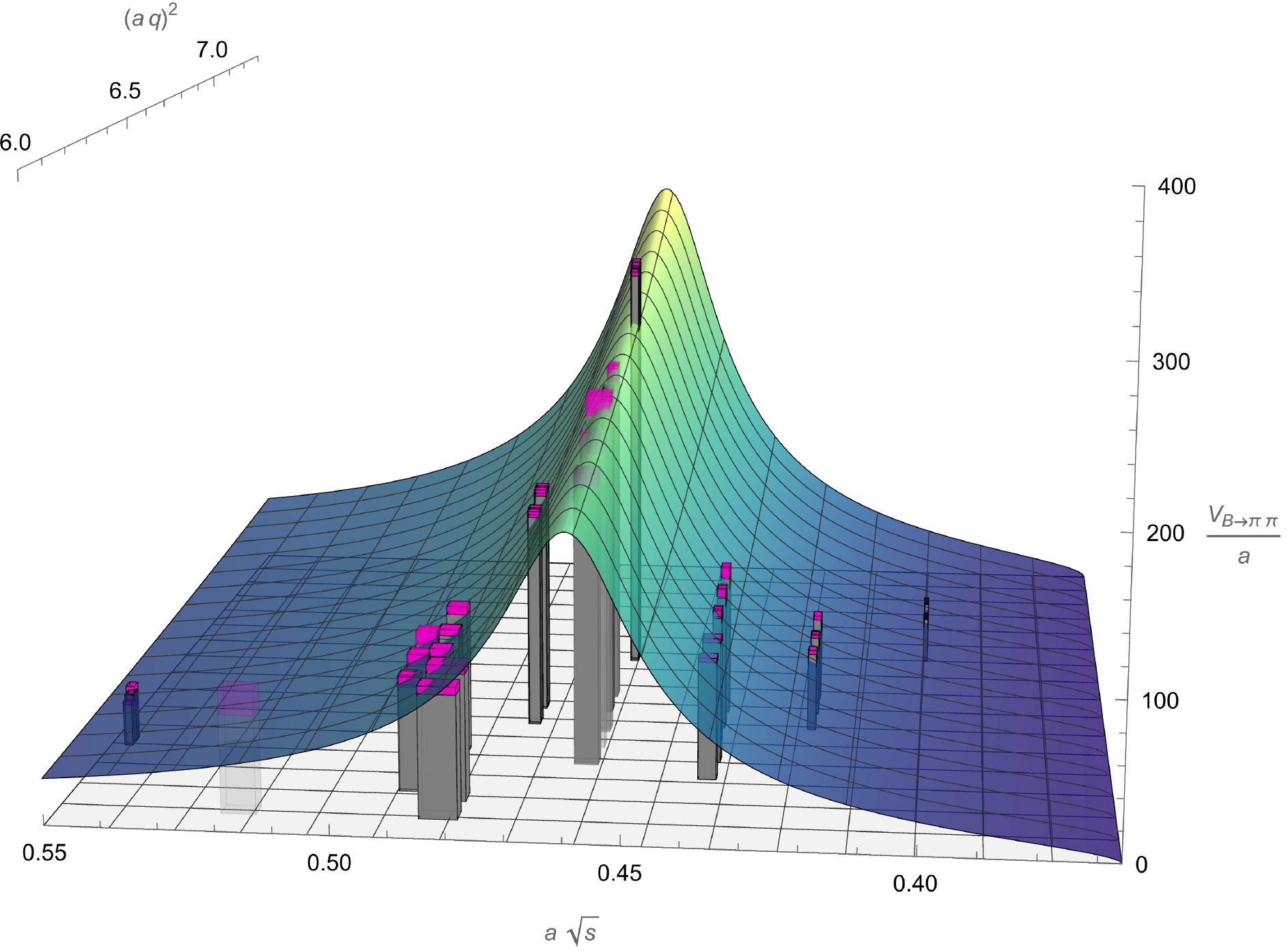}
    \caption{A 3-D plot for the transition amplitude; shown is the scattering amplitude BWII + N1M1, while the histogram points are the data with statistical uncertainties only.}
    \label{fig:3D}
\end{figure}

To elaborate on the analysis, we must first discuss the transition amplitude that enters the matrix element in the infinite volume. The infinite-volume matrix element obeys a Lorentz decomposition, which for the vector current of the $B\to\pi\pi\ell\bar{\nu}$ looks like:
\begin{align}
    \langle n | J_\mu | B \rangle_\infty = \frac{2 i V(E^{\star},q^2)}{m_B + 2m_\pi} \epsilon_{\mu\nu\alpha\beta} \, \varepsilon^{\nu*} \, P^{\alpha} p_{B}^{\beta},
\end{align}
where $P$ is the final $\pi\pi$ state four-vector,  $\varepsilon$ is the polarization four-vector of the $\pi\pi$ state in $P$-wave and $p_B$ is the initial state $B$-meson four vector. The momentum transfer vector $q_\mu = P_\mu - (p_B)_\mu$ and the momentum transfer is $q^2 = q_\mu q^\mu$; $\epsilon$ is the $4$-d Levi-Civita tensor. The quantity $V(E^{\star},q^2)$ is the transition amplitude and can be written down in many ways - however, there are a few requirements: unitarity and analyticity. Both are satisfied if we write $V$ as:
\begin{align}
    V(E^{\star},q^2) = F(E^{\star},q^2) \frac{T(E^{\star})}{k},
\end{align}
where $k$ is the $\pi\pi$ scattering momentum, $T$ is the scattering amplitude and $F$ is new function which we will call the form-factor. For the scattering amplitude $T$ we use the parameterizations from Ref.~\cite{Alexandrou:2017mpi}, and for the form factor $F$ we use a generalized $z$-expansion \cite{Alexandrou:2018jbt}:
\begin{align}
\label{eq:zexp}
    F(E^\star, q^2) = \frac{1}{1-\frac{q^2}{m_P^2}} \sum_{n,m} A_{n,m} \, z^{n}(q^2) \, (E^{\star 2} - E_{thr}^{2})^{m},
\end{align}
where $E_{thr}=2m_\pi$. To relate the infinite-volume matrix element with the finite-volume matrix element, we recall the two-point correlation function in a finite-volume being expanded as a sum of poles, where the poles are located at the finite-volume energies. Nevertheless, an expansion over poles involves the pole location, $E_n^\star$, and the pole residuum $R_n$. While the location describes the finite-volume energies, the residue $R_n$ describes the normalization of the state belonging to that pole:
\begin{align}
\label{eq:FVml}
    \langle n | J_\mu | B \rangle_L = \sqrt{R_n} \langle n | J_\mu | B \rangle_\infty.
\end{align}
The residue $R_n$ is defined as:
\begin{align}
R_n \, = \lim_{E^\star \to E_n^\star} \frac{E^\star-E_n^\star}{F^{-1}+T},
\end{align}
while the details of the best way to calculate the residue are discussed in Ref.~\cite{Briceno:2021xlc}. To fit the transition amplitude to the lattice data, we thus pick a model for the infinite-volume transition amplitude and calculate the finite-volume matrix element according to Eq.~\ref{eq:FVml}. We build a $\chi^2$ function from there, which we minimize. For the model transitions amplitudes, we use two types of the scattering amplitudes $T$, $BWI$ - a simple Breit-Wigner, and $BWII$ - a simple Breit-Wigner modified by Blatt-Weisskopf \cite{VonHippel:1972fg} barrier factors. For the form factors $F$, we use $4$ models, that truncate the $z$-expansion at $n=1$ and $m=1$ as listed in Tab.~\ref{tab:Fs}
\begin{table}[htb!]
    \centering
    \begin{tabular}{|c|c|c|c|}
    \hline
    model  & $n_{max}$ & $m_{max}$ & $\chi^2/{\rm dof}$ \\
    \hline
    BWI + N0M0   & $0$       & $0$       & $1.18$ \\
    BWI + N1M0   & $1$       & $0$       & $1.07$ \\
    BWI + N0M1   & $0$       & $1$       & $0.63$ \\
    BWI + N0M0   & $1$       & $1$       & $0.50$ \\
    \hline
    BWII + N0M0   & $0$       & $0$       & $0.55$ \\
    BWII + N1M0   & $1$       & $0$       & $0.55$ \\
    BWII + N0M1   & $0$       & $1$       & $0.52$ \\
    BWII + N0M0   & $1$       & $1$       & $0.51$ \\
    \hline
    \end{tabular}
    \caption{Results of fits to $8$ different models of the $B\to\pi\pi \ell\bar{\nu}$ transition amplitude. The left column lists the model for the scattering amplitude + the model for the form factor. In the middle two columns, the truncation limits of the sum in Eq.~\ref{eq:zexp} are listed, and the right-most column shows the $\chi^2/{\rm dof}$.}
    \label{tab:Fs}
\end{table}
The average of all the $F$'s combined is shown in Fig.~\ref{fig:ffx}, where the dark-shaded region represents the statistical uncertainty, while the light-shaded region represents the systematical uncertainty. The 3-D representation is shown in Fig.~\ref{fig:3D} for the transition amplitude BWII + N1M1, where the curve shows the fitted transition amplitude. At the same time, the rectangles represent the data with statistical uncertainties only (mapped into the infinite volume).

\section{Prospects for the Near Future} \label{sec:4}

The prospects in applying the formalism discussed here span Electroweak physics and Photoproduction. While photoproduction processes might not initially appear central to high-energy and nuclear physics, they offer crucial insights into the least understood aspects of the Standard Model, particularly the structure and contents of hadronic resonances. The ongoing debate over the nature of various hadronic resonances, such as the $a_0(980)$, $f_0(980)$, $D_0^\star(2400)$, and $D_{s0}^\star(2317)$, extends beyond just their masses and decay widths. Comprehensive understanding requires additional data, where lattice QCD, supported by experimental evidence, can contribute significantly. This approach aims to foster a chemistry-like comprehension of hadrons underpinned by theoretical predictions, experimental validations, and intuitive models. Lattice QCD can play a pivotal role in determining radiative transitions in these states, guiding experimental measurements, and extending our knowledge to yet unmeasured quantities such as form factors and other structure observables \cite{Briceno:2019nns,Briceno:2020xxs,Briceno:2020vgp} of resonances.

Numerous experimentally observed processes in Flavor physics await further theoretical backing from lattice QCD. Notable among these are the $b\to s \ell \ell$ transitions, extensively measured by Belle II, LHCb, CMS, and ATLAS \cite{Belle-II:2022fky,LHCb:2020gog,CMS:2020oqb,ATLAS:2018gqc}. Key processes include $B\to K^\star(892) \ell \ell$ and $B\to K_0^\star(700)\ell \ell$, along with transitions linked to the determination of $V_{ub}$, such as $B\to \rho \ell \bar{\nu}$ and $B\to f_0(500) \ell \bar{\nu}$. Their charm quark analogs, like $D \to \rho\ell\bar{\nu}$ and $D \to K^\star(892) \ell\bar{\nu}$, also present opportunities for precision tests that could reveal new physics. The $B \to D \pi \ell \bar{\nu}$ process, offering insights into the $B\to D^{\star\star} \ell\bar{\nu}$ puzzle \cite{Gustafson:2023lrz} and the inclusive vs exclusive debate over $V_cb$, is another area where lattice QCD could significantly contribute. Due to the very narrow nature of the $D^\star$ meson, the finite-volume effects in $B\to D^\star \ell \bar{\nu}$ are tiny (sub percent according to limits in App. B of Ref.~\cite{Briceno:2015csa}).

While each mesonic process outlined above has a partner baryon process, the situation is different. The processes related to the $b \to s \ell \ell$ have undergone a first iteration \cite{Meinel:2020owd,Meinel:2021rbm,Meinel:2021mdj,Farrell:2023vnm}, and improvements are planned \cite{Meinel:2023wyg}; however, some of the final states are hadronically unstable and will inevitably require the utilization of the finite-volume formalism in the analysis. This poses a challenge, considering spectroscopy studies in the baryon sector lag behind those in the meson sector. On the topic of nucleon transitions, there are essential and exciting processes related to the $\Delta(1232)$. At first, it would be great to see a similar program for the $\Delta(1232)$ as was seen for the $\rho$ resonance; this would provide us with sufficient confidence to determine also the transition amplitudes for $N\gamma \to \Delta(1232)$ and the $N (V-A) \to \Delta(1232)$, which will play a crucial role in the event simulation and identification at DUNE \cite{NuSTEC:2017hzk}.

\section{Summary} \label{sec:Summary}
The study of electroweak transitions is crucial for understanding particle behavior, especially in processes altering quark flavor or lepton number. These transitions not only enrich our understanding of the Standard Model, particularly in aspects like the Standard Model parameters, but also explore CP violation and the matter-antimatter asymmetry of the universe and use rare decays as probes for Physics Beyond the Standard Model. Lattice Quantum Chromodynamics (QCD) has been vital in determining the elements of the CKM matrix thus far. However, a gap remains in the study of hadronic resonances, opening an opportunity for many new processes.\\

The finite-volume lattice QCD methodology involves complex computational techniques to determine infinite-volume transition amplitudes, encompassing spectrum and matrix element analysis. However, both aspects have matured sufficiently that any collaboration can pick them up and utilize them in phenomenological calculations of their choice. Furthermore, to those who still need more confidence, try, and if you get stuck, know that you are in an open community that always welcomes questions.

\section*{Acknowledgments}

I am grateful to my collaborators Stefan Meinel, Marcus Petschlies, John W. Negele, Srijit Paul and Andrew Pochinsky. We thank Kostas Orginos, Balint Joó, Robert Edwards, and their collaborators for providing the gauge-field configurations. Computations for this work were carried out in part on (1) facilities of the USQCD Collaboration, which are funded by the Office of Science of the U.S.~Department of Energy, (2) facilities of the Leibniz Supercomputing Centre, which is funded by the Gauss Centre for Supercomputing,  (3) facilities at the National Energy Research Scientific Computing Center, a DOE Office of Science User Facility supported by the Office of Science of the U.S.~Department of Energy under Contract No.~DE-AC02-05CH1123, (4) facilities of the Extreme Science and Engineering Discovery Environment (XSEDE), which was supported by National Science Foundation grant number ACI-1548562, and (5) the Oak Ridge Leadership Computing Facility, which is a DOE Office of Science User Facility supported under Contract DE-AC05-00OR22725. L.L. acknowledges the project (J1-3034) was financially supported by the Slovenian Research Agency.

\bibliographystyle{utphys-noitalics}
\bibliography{pos}

\end{document}